\documentstyle[prl,aps,multicol,epsf]{revtex}

\begin{document}

\draft

\title{Persistent Orbital Degeneracy in Carbon Nanotubes}

\author{A. Makarovski,$^{1}$ L. An,$^{2}$ J. Liu,$^{2}$ and G. Finkelstein$^{1}$}

\address{ $^1$ Departments of Physics, Duke University, Durham, NC 27708}
 
\address{ $^2$ Department of Chemistry, Duke University, Durham, NC 27708}

\maketitle

\begin{abstract}
The quantum-mechanical orbitals in carbon nanotubes are doubly degenerate over a large number of states in the Coulomb blockade regime. We argue that this experimental observation indicates that electrons are reflected without mode mixing at the nanotube-metal contacts. Two electrons occupying a pair of degenerate orbitals (a ``{\it shell}'') are found to form a triplet state starting from zero magnetic field. Finally, we observe unexpected low-energy excitations at complete filling of a four-electron shell.

\end{abstract}


\begin{multicols}{2}
\section{Introduction}

Single-wall carbon nanotubes exhibit a variety of interesting transport phenomena at low temperatures \cite{CNTreview}.  At weak coupling to the metal electrodes the whole length of the nanotube may serve as single Quantum Dot, demonstrating a Coulomb blockade behavior at low temperatures \cite{Tans97,Bockrath97}. As the quality of the nanotube samples have improved over time, the single-electron conductance peaks in the Coulomb blockade regime have been found to group in clusters of four \cite{Liang2002,Buitelaar2002,Cobden2002,Babic,Moriyama,Sapmaz,Jarillo-Herrero}. This behavior has been attributed to pairs of quantum-mechanical orbitals that originate in two subbands of the nanotube electronic dispersion and are close in energy, forming four-electron ``shells''.

In this work, we investigate the shell structure in metallic carbon nanotubes. We find that in a large percentage of shells the two orbitals are degenerate with a possible level mismatch not exceeding one tenth of the shell spacing. The existence of degeneracy in real nanotube devices does not follow trivially from the symmetry properties of 2D graphite, as claimed in Refs. \cite{Buitelaar2002,Cobden2002,Babic,Moriyama}. Indeed, a sizable orbital mismatch was found in most studies \cite{Liang2002,Buitelaar2002,Cobden2002,Babic,Moriyama,Sapmaz,Jarillo-Herrero}. In this paper, we report on experimental observation of the orbital degeneracy over a large number of states. Our finding also demonstrates that the electrons are reflected without mode mixing at the metal contacts. By studying magnetic field dependence, we find that two electrons occupying a pair of degenerate orbitals form a triplet state starting from zero magnetic field. Finally, we observe unexpected low-energy excitations when a four-electron shell is completely filled, both in the sequential tunneling regime and in the inelastic cotunneling. 

The nanotubes were grown by a CVD method using CO as a feedstock gas (the details are described in our earlier publication \cite{Zheng2002}), and Cr/Au contacts were patterned by e-beam lithography. 
We present results measured on metallic nanotubes. We measure the differential conductance at temperatures of 0.3 or 1.2 K by a standard AC technique at the excitation level of 10-50 $\mu$V RMS.

\section{Level Degeneracy}

\begin{figure}
\epsfxsize=\linewidth
\epsfbox{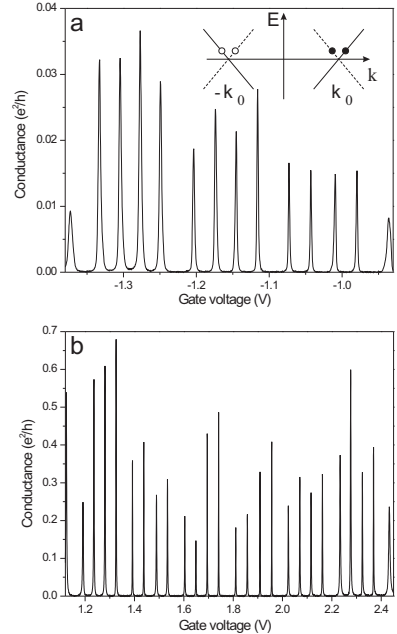}
\caption{\narrowtext
a) Differential conductance of a 750 nm-long nanotube as a function of $V_{gate}$ measured at the temperature of 1.2 K. Inset: schematics of the band structure in a metallic nanotube. b) Differential conductance of a 600 nm-long nanotube as a function of $V_{gate}$ measured at the temperature of 0.3 K.
}
\end{figure}

Figure 1a shows the Coulomb blockaded conductance measured in a narrow range of gate voltages on a 750 nm long nanotube at $T= 1.2$ K. Most single-electron conductance peaks cluster in groups of four, where the peaks within each cluster have similar heights \cite{Liang2002,Buitelaar2002,Cobden2002,Babic,Moriyama,Sapmaz,Jarillo-Herrero}. The spacing between the peaks in each cluster is smaller than the spacing between the neighboring clusters. Figure 1b demonstrates a similar peak clustering measured on a 600 nm long nanotube at $T= 0.3$ K.
This behavior is further illustrated in Figure 2, where we plot the {\it addition energies} required to add consecutive electrons to the nanotube of Figure 1a. (The ``addition energy'' \cite{QDreview} is measured as the gate voltage spacing between the neighboring conductance peaks multiplied by a factor $\alpha \equiv C_{gate}/C_{total}$ which we later find to be $\alpha=0.13$ from Figure 4.)

The exceptional reproducibility of the pattern over a wide range of gate voltages and the relatively large nanotube length allows us to trace the positions of hundreds of conductance peaks within the same sample. We extract the peak positions from the successive gate voltage sweeps and keep only the peak positions which coincide in different sweeps. By doing this, we get rid of the relatively rare random offset charge events, which rigidly shift the conductance curves in narrow ranges of gate voltage. One can see extended ranges of gate voltage where three successive addition energies are very close and each fourth addition energy is significantly ($\sim 50 \%$) larger than the base line (one of such regions is illustrated in Figure 2b). This pattern follows from the peak clustering observed in Figure 1, where the larger addition energies correspond to the spacing between the clusters. 

\begin{figure}
\epsfxsize=\linewidth
\epsfbox{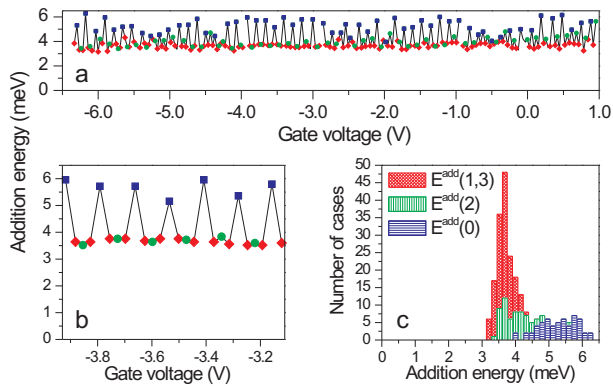}
\caption{\narrowtext
a) Addition energies (spacings between the neighboring conductance peaks multiplied by $\alpha$) as a function of $V_{gate}$, measured on the same sample as in Figure 1a. b) Same in a narrower range of $V_{gate}$. Blue squares: addition energies corresponding to a complete shell filling ($E^{add}_0$); green circles: addition energy corresponding to a half-filled shell ($E^{add}_2$); red diamonds: addition energies of the first and the third electrons in a shell ($E^{add}_1$, $E^{add}_3$). $E^{add}_0$ points are noticeably higher than the base line. c) Histogram of the addition energies.
}
\end{figure}

Let us for now assume the ``Constant Interaction Model'' \cite{QDreview,ABG} and ignore the  mismatch $\delta$ between the orbitals in a shell. Then, a constant charging energy $E_c$ is required to add an electron to a partially filled shell. A larger energy $E_c + \Delta$, where $\Delta$ is the shell spacing, is required to add an electron to the next shell. As a result, the addition energies within each shell are equal to $E_c$, while the addition energy required to put an electron to a new shell is $E_c + \Delta$, in agreement with observations in Figures 2a and b. Experimentally, we obtain $E_c \approx $ 3.5 meV and $\Delta \approx$ 2 meV. The latter agrees well with the expected $\Delta = \pi \hbar v_F / L$ (=2 meV for $L= 750$ nm). Such a simple description may be applicable because the exchange interaction $J$ and the excess interaction of two electrons occupying the same orbital $\delta U$ are expected to be small compared to the shell spacing $\Delta$, as discussed in the next section (see also Appendix) \cite{Oreg2000,Ke2003}. Both $J$ and $\delta U$ are further reduced by the relatively large diameter of our nanotubes ($\sim 3 nm$). 

The regular four-electron shell filling is disrupted at some gate voltages, where electrons occupy non-degenerate orbitals. In such cases, one observes pairs of conductance peaks, closely spaced and of similar heights \cite{Cobden2002}. The addition energy of the second electron in a pair is smaller than the addition energy of the first electron (``even-odd oscillations''), as visible in Figure 2a {\it e.g.} around $V_{gate}=-4.5$ V. It is important, that while the orbital degeneracy is absent at some gate voltages, it is observed over other extended ranges of gate voltage. 

Apparent clustering of four conductance peaks does not yet guarantee the orbital degeneracy. Indeed, the clusters are visible even when the two orbitals are not degenerate, and the level mismatch $\delta$ could be as high as 0.2 to 0.4 of $\Delta$ \cite{Liang2002,Buitelaar2002,Moriyama,Sapmaz}.
Also, a relatively large $\delta$ may be masked by the high contact transparency that introduces lifetime broadening of the peaks \cite{Liang2002,Babic,Jarillo-Herrero}. In our measurements, the peaks are narrow and the orbital degeneracy is revealed: we find, that for a large fraction of the shells $\delta$ is negligible. For example, in Figure 2b, the largest level mismatch $\delta$ is observed at $V_{gate}= -3.34$ V (the second addition energy in a shell is slightly larger than its neighbors). Here $\delta=$ 0.2 meV is equal to 0.1 $\Delta$. The low values of the orbital mismatch, observed over a large number of states, point to the existence of a mechanism establishing the orbital degeneracy, rather than an accidental alignment of levels. 

The orbital degeneracy survives significant variations in $\Delta$ that are likely caused by the disorder potential (notice variations of the inter-cluster spacings in Figure 2b). It is known that long-range potential does not effectively introduce back scattering and mode mixing in metallic nanotubes \cite{Ando2000,McEuen99}. At the same time, the orbital degeneracy can be easily lifted by structural imperfections in the nanotube, such as pentagon-heptagon pairs \cite{Ke2003}, which introduce back scattering and mode mixing \cite{Ando2000}. Apparently, these structural defects are relatively rare in our samples. 

Experimental observation of the orbital degeneracy makes a non-trivial statement regarding the properties of the contacts: evidently, reflections at the contacts do not mix the subbands. Let us consider these reflections in more details. We surmise that one of the degenerate orbitals originates in the vicinity of point $k_0$ in the nanotube dispersion (inset of Figure 1) and another orbital originates in the vicinity of point $-k_0$ \cite{Saito}. This means than an electron in state $k_0 + \delta k$ and moving to the right is reflected from the contact to the state at $k_0 - \delta k$ (solid circles in the inset of Figure 1), and not to the state $-k_0 - \delta k$. 
Similarly for the other orbital, electron in $-k_0 + \delta k$ is reflected solely to $-k_0 - \delta k$ (open circles in the inset of Figure 1). Such processes require an electron to change the dispersion branch upon reflection (solid and dashed lines in the inset of Figure 1). In fact, Ref. \cite{Kong1999} suggests that the metal contact perturbs the energy spectrum of the nanotube, mixing the two branches and opening a gap around $\pm k_0$. The resulting spectrum {\it at the contacts} then look like that of a semiconductor, and the electron could adiabatically move from $k_0+ \delta k$ to $k_0- \delta k$ upon reflection. 

An alternative picture (see {\it e.g.} a sketch in Figure 1a of \cite{Sapmaz}), where the states at $k_0 + \delta k$ and $k_0 - \delta k$ belong to different orbitals, seems unlikely in our case because (i) reflection from $k_0 + \delta k$ to $- k_0 - \delta k$ requires a large momentum transfer $\sim 2k_0$, {\it i.e.} a sharp nanotube-metal interface. If so, the scattering to $k_0 - \delta k$ would be significant, and the two orbitals would mix and split in energy, contrary to our observations. (ii) Any possible degeneracy between the orbitals made of the states [$k_0 + \delta k$,  $-k_0 - \delta k$] and [$k_0 - \delta k$, $-k_0 + \delta k$] would be only accidental. Indeed, in this case the ``particle in a box'' quantization conditions yield for the two types of orbitals the energies of $\hbar v_F ({\pi N_1 \over L} - k_0)$ and $\hbar v_F (k_0 - {\pi N_2 \over L} )$, where $N_{1,2}$ are integers. The two sets of states coincide in energy only if $2 k_0 L / \pi$ equals an integer, a condition that cannot be satisfied for a majority of states in a real sample. 

Finally, we analyze the {\it statistics} of the addition energies (for reviews on this subject see \cite{ABG,Alhassid2000}). We divide the addition energies into three categories: the addition energy corresponding to a complete shell filling ($E^{add}_0 = E_c + \Delta$ blue squares in Figure 2b), the addition energy corresponding to a half-filled shell  ($E^{add}_2=E_c + \delta$, green circles), and the addition energies of the first and the third electrons in a shell ($E^{add}_1=E^{add}_3= E_c$, red diamonds). The statistical distribution of these energies is plotted by the corresponding colors in Figure 2c. As expected, the overall distribution is clearly bimodal, where the main lobe corresponds to the baseline of Figure 2a, {\it i.e.} to the charging energy. The high energy shoulder of the distribution is comprised of $E_c + \delta$ and $E_c + \Delta$. The distribution of $\delta$ has a significant weight at $\delta=0$ and for more than $40\%$ of states $\delta \lesssim 0.1 \Delta$. We conclude that in a significant fraction of the shells the level repulsion is suppressed, and the level degeneracy is {\it not accidental}. Finally, it appears that the distribution of $\Delta$ is broader than the distribution for $E_c$ (as seen in Figures 2b and c).

\section{Behavior in Magnetic Field}

\begin{figure}
\epsfxsize=\linewidth
\epsfbox{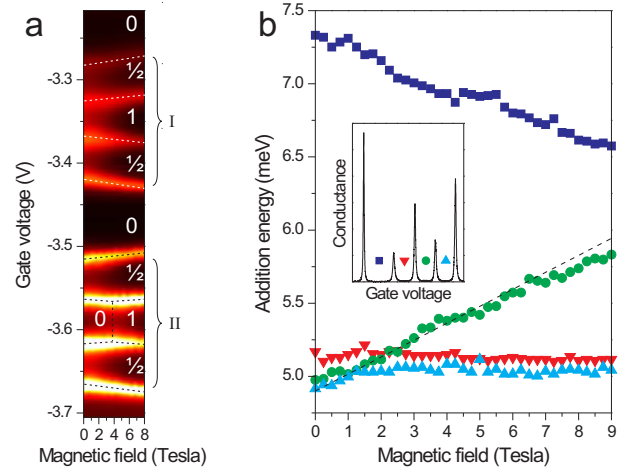}
\caption{\narrowtext
a) The nanotube conductance as a function of $V_{gate}$ and $B_{\perp}$ measured on a 600 nm long nanotube. The conductance is color coded (bright colors correspond to a higher conductance, dark colors correspond to a lower conductance). The inferred spin sequence is indicated by numbers overlaying the image. For cluster II, the two middle traces exhibit a kink, indicating a singlet-triplet transition.
b) Addition energies for the four electrons in a shell (similar to shell I in Figure 3a) measured as a function of $B_{\perp}$ on a different 600 nm long nanotube. $E^{add}_1$ and $E^{add}_3$ stay remarkably constant in magnetic field, indicating that two electrons in a row (1st and 2nd; 3rd and 4th) are added with the same spin projections. The addition energy $E^{add}_2$ corresponding to a half-filled shell grows in $B_{\perp}$ with a slope close to $2\mu_0$ (dashed line). Inset: conductance plot from which the addition energies where extracted. The symbols correspond to those in the main panel. 
}
\end{figure}

Figure 3a shows the nanotube conductance as a function of the gate voltage and magnetic field perpendicular to the nanotube axis, $B_{\perp}$, measured on a 600 nm long nanotube. At zero field, we see that the peaks form two clusters (I and II) of four electrons, similar to Figure 1. The perpendicular field couples primarily to the spin of the electrons in the nanotube and not to their orbital motion, which allows us to study the spin state of the system. In cluster I, we observe that the two bottom  traces move down in magnetic field, and the two top traces move up. This behavior indicates that pairs of consecutive electrons are added to the nanotube with the same spin projection (i.e. spin up or spin down). Therefore, two electrons occupying a shell form a triplet state \cite{Liang2002,Jarillo-Herrero}, so that the sequence of the total spin of the nanotube as the electrons fill the shell is $0-1/2-1-1/2-0$. Such a behavior is in contrast with alternating $0$ and $1/2$  states found at zero field in shells with non-degenerate levels \cite{Buitelaar2002,Cobden2002,Moriyama}.

In cluster II, the second and the third single-electron traces exhibit kinks around $4$ Tesla, a behavior observed in \cite{Buitelaar2002,Moriyama}. Namely, at low $B_{\perp}$, the second trace from the bottom shifts up and the third trace shifts down with the field, indicating that the four electrons in shell II enter the nanotube with alternating spin projections. In higher fields, these directions change: the second peak shifts down and the third peak shifts up, just as the traces in cluster I. Evidently, in shell II a small level mismatch exists at zero field. As a result, at fields below 4 Tesla, the two electrons form a singlet state on the lower orbital, while at fields above it, the Zeeman energy forces the two electrons into a triplet state. From the value of the field at the kink location, we find the level mismatch in shell II of $\delta \sim 0.5$ meV (relatively small, compared to the shell spacing $\Delta \approx 3$ meV in this sample). Any possible level mismatch in shell I is at least several times smaller.   

To further quantify the observed behavior, in Figure 3b we present the addition energies measured as a function of $B_{\perp}$ on a different 600 nm nanotube in a shell similar to shell I in Figure 3a. The cluster of four peaks where the data were measured is shown in the inset. The four different symbols mark the valleys from which the four addition energies ($E^{add}_{0-3}$) were extracted for the main panel of Figure 3b. We see that $E^{add}_1$ and $E^{add}_3$ (triangles) stay remarkably constant in magnetic field. At the same time, $E^{add}_2$ (circles) grows steadily with magnetic field with a slope close to $2\mu_0$ (dashed line; $g$-factor in nanotubes is expected to be close to 2). These observations prove that the first and the second electron are added with the same spin projection (spin-up), while the third and the fourth electron are both added spin-down. In particular, this supports our statement that the two-electron ground state is a triplet down to the lowest magnetic fields.  Finally, $E^{add}_0$ (squares) is significantly larger (by approximately $\Delta$) than $E^{add}_{1-3}$ and steadily decreases at a slope close to $2\mu_0$, as expected. 

For degenerate levels ($\delta=0$), the exchange interaction $J$ and excess interaction $\delta U$ should show up in $E^{add}_2$ being greater than $E^{add}_{1,3}$ at zero magnetic field (see Appendix). From Figure 3b we have to conclude that the possible $J$ and $\delta U$ are very small, not exceeding $0.1 \Delta$ \cite{Oreg2000,Ke2003}. It is also noticeable from Figure 3b that $E^{add}_1>E^{add}_3$. (Depending on the particular shell, we find that this inequality may be reversed, or the two energies may be equal.) This behavior indicates that two electrons occupying the first (slightly lower in energy) orbital in a shell have a stronger repulsion than the two electrons occupying the second orbital (see Appendix). 

\section{Low-Energy Excitations}
Let us turn to the nanotube conductance at relatively large source-drain voltages $V_{sd}$. Figure 4a shows the conductance map as a function of the gate voltage $V_{gate}$ and $V_{sd}$. The dark regions of suppressed conductance along the horizontal axis (``Coulomb diamonds'') indicate that electron transport is blocked and the number of electrons in the nanotube is fixed. The sizes of the Coulomb diamonds follow the same regular pattern as in Figure 1: three relatively small diamonds of a similar size follow a larger diamond. 

\begin{figure}
\epsfxsize=\linewidth
\epsfbox{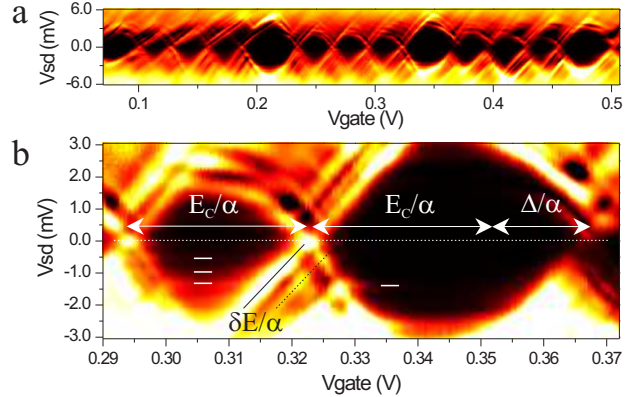}
\caption{\narrowtext
a) Differential conductance map as a function of $V_{gate}$ (horizontal axis) and the source-drain bias (vertical axis). Each fourth Coulomb diamond is clearly larger than its neighbors. b) Zoomed-in view of the $4N$ and $4N-1$ - electron diamonds. A low-lying
excitation of the $4N$ electron state is shown by the dashed line. Its energy $\delta E$ is clearly smaller than the shell spacing $\Delta$. White horizontal bars indicate the co-tunneling thresholds.
}
\end{figure}Figure 4b shows the zoomed-in view of the $4N-1$ and $4N$ electron diamonds in Figure 4a. The lines bordering the diamonds correspond to the resonances between the energy level in the nanotube and the Fermi energy of either lead. We use them to extract the coefficient $\alpha$ \cite{QDreview}. Similarly, the inclined lines running parallel  to the diamond boundaries correspond to the resonances between the Fermi energy at the contacts and {\it excited} states of the nanotube in the sequential tunneling regime. The horizontal lines visible inside the Coulomb diamonds (marked by white dashes in Figure 4b) correspond to the inelastic cotunneling events, where an excitation is created as a result of an electron tunneling through the nanotube \cite{Averin1992}. 

The states marked by the inclined solid and dashed lines in Figure 4b correspond to the ground and the excited states of the nanotube with $4N$ electrons (the topmost occupied four-electron shell is completely full). We find that these states are separated by an energy $\delta E$ of only $\sim 0.7$ meV, several times smaller than the shell splitting $\Delta$ extracted the ground state energies in Figure 2 (the scale of $\Delta$ can also be gauged from the size difference between the $4N$ and $4N-1$ diamonds in Figure 4). This observation is especially surprising, given that the ground state with a completely filled topmost four-electron shell should be separated by $\sim \Delta$ from the lowest excited state. (This is true up to corrections due to $J$, $\delta U$ and $\delta$, which in our case are negligible compared to $\Delta$, see discussion in the preceding sections.) 
The energy of the smallest excitation measured through the inelastic cotunneling threshold inside a $4N$-electron diamond (not shown) coincides with $\delta E$ extracted in the sequential tunneling regime.

We consider several possible mechanisms to account for $\delta E$: quantized 1D vibrational modes of the nanotube, features in the density of states of the leads, or collective electronic excitations: ({\it i}) $\delta E$ probably does not originate in the quantized phonon modes of the nanotube, as it would require a relatively long segment of the nanotube (tens of nanometers) to be decoupled from the substrate. ({\it ii}) The same $\delta E$ is visible in sequential tunneling in resonance with both the source and drain electrodes. It seems unlikely that both leads would possess a similar feature in the density of states. ({\it iii}) Alternatively, $\delta E$ may represent an energy of a collective  electronic excitation. It was recently suggested that in 1D GaAs channels at low densities the interactions may renormalize the energies of the spin excitations to lower values \cite{Matveev2004}. In our experiment, the density of carriers near the contacts may possibly be reduced by opening of the local semiconducting gap as described above in Section II. In this case, the low-density 1D gas near the contacts is strongly correlated and may sustain the low energy excitations $\delta E$ similar to the prediction of Ref. \cite{Matveev2004}. At the same time, the single-particle energy $\Delta$ would be determined by the properties of the nanotube ``bulk'', where interactions are relatively less important. Further work is required to identify the origin of the low-energy excitations $\delta E$.

\section{Conclusions}

We investigate the degeneracy of orbitals in metallic carbon nanotubes. We argue that the degeneracy does not follow directly from the presence of two subbands in the electronic dispersion of a nanotube. Indeed, a sizable orbital mismatch was found within a shell in prior studies \cite{Liang2002,Buitelaar2002,Cobden2002,Babic,Moriyama,Sapmaz,Jarillo-Herrero}. We present an explanation of the degeneracy and discuss the necessary conditions for its observation. Two electrons filling a pair of degenerate levels are found to form a triplet state down to lowest magnetic fields. Finally, when a four-electron shell is completely filled, the large and clearly identifiable single-particle gap allows us to observe unexpected low-energy excitations, which may serve as an indication of collective electronic modes.

\acknowledgements
We thank I. Aleiner, H. Baranger, R. Berkovits, A. Chang, B. Halperin, K. Matveev, Y. Oreg and D. Ullmo for valuable discussions and M. Prior and A. Zhukov for technical assistance. The work is supported by NSF DMR-0239748.

\appendix
\section{}
\label{sec:appendix}

In our analysis we use an effective Hamiltonian in the spirit of Ref. \cite{Oreg2000}. The expression invariant under rotations of the total spin is derived starting from the Hartree-Fock Hamiltonian and using operator identities of Ref. \cite{Kurland2000}. The resultant expression depends on ${\it N_i}$ (0, 1 or 2), the occupation number of orbital $i$, and on ${\it {\bf S}_i}$ (0, $\frac{1}{2}$ or 0, respectively), the total spin of the electrons occupying this orbital:

\begin{eqnarray} 
&&
E([N_i, S_i])= 
\sum_i \epsilon_i N_i + \frac{1}{2} \sum_{i \not= j} U_{ij} N_i N_j
\nonumber \\
&&
\hphantom{[E}
+\frac{1}{2} \sum_{i} U_{ii} N_i (N_i-1)-
\sum_{i \not= j} J_{ij} (\hat{\bf S}_i \cdot \hat{\bf S}_j + \frac{1}{4}  N_i N_j)
\label{eq:E1}
\end{eqnarray}
Here $\epsilon_i$ is the bare energy of an electron occupying orbital $i$; $U_{ij}$ and $J_{ij}$ are respectively the direct and exchange interactions between two electrons on orbitals $i$ and $j$. For the practically relevant case of two topmost (partially) occupied orbitals $\alpha$ and $\beta$ the expression reduces to 

\begin{eqnarray} 
&&
E(N_\alpha, N_\beta, S_{tot})= 
\sum_{i=\alpha,\beta} \epsilon_i N_i + U_{\alpha \beta} N_\alpha N_\beta
\nonumber \\
&&
\hphantom{E(N}
+\frac{1}{2}  \sum_{i=\alpha,\beta} U_{ii} N_i (N_i-1)
\nonumber \\
&&
\hphantom{E(N}
-J_{\alpha \beta} (S_{tot}^2 - S_{\alpha}^2 - S_{\beta}^2+ \frac{1}{2} N_\alpha N_\beta)
\label{eq:E2}
\end{eqnarray}
where $S_{tot}$ is the total spin of the electron system, Here we disregarded interactions with the lower, completely filled orbitals. The energies of the relevant states with different electronic occupations (presuming that orbital $\alpha$ is slightly lower than $\beta$) are:

$E(1,0,\frac{1}{2})= \epsilon_{\alpha}$ (1e),

$E(2,0,0)=2 \epsilon_{\alpha}+U_{\alpha \alpha}$ (2e singlet),

$E(1,1,1)=\epsilon_{\alpha}+ \epsilon_{\beta}+U_{\alpha \beta}-J_{\alpha \beta}$ (2e triplet),

$E(1,1,0)=\epsilon_{\alpha}+ \epsilon_{\beta}+U_{\alpha \beta}+J_{\alpha \beta}$ (2e singlet),

$E(2,1,\frac{1}{2})=2\epsilon_{\alpha}+ \epsilon_{\beta}+2U_{\alpha \beta}+U_{\alpha \alpha}-J_{\alpha \beta}$ (3e),

$E(2,2,0)=2\epsilon_{\alpha}+ 2\epsilon_{\beta}+4U_{\alpha \beta}+U_{\alpha \alpha}+U_{\beta \beta}-2J_{\alpha \beta}$ (4e).

The addition energies measured in experiment are second differences of these $E(N_\alpha, N_\beta, S_{total})$. If the ground state of two electrons is a triplet, 

$E^{add}_1=\epsilon_{\beta}-\epsilon_{\alpha}+U_{\alpha \beta}-J_{\alpha \beta}$,

$E^{add}_2=\epsilon_{\alpha}-\epsilon_{\beta}+ U_{\alpha \alpha} + J_{\alpha \beta}$, and

$E^{add}_3=\epsilon_{\beta}-\epsilon_{\alpha}+U_{\alpha \beta}+U_{\beta \beta}-U_{\alpha \alpha}-J_{\alpha \beta}$. 

If we assume $U_{\alpha \alpha}=U_{\beta \beta}$ and define $\delta \equiv \epsilon_{\beta}-\epsilon_{\alpha}$ (level mismatch), $U \equiv U_{\alpha \beta}$ and  $\delta U \equiv U_{\alpha \alpha}- U$ (excess interaction), the expressions for addition energies become 

$E^{add}_1=E^{add}_3= U-J+\delta $, and 

$E^{add}_2= U+\delta U +J -\delta$.

\end{multicols}

\begin{references}

\bibitem{CNTreview} C. Dekker, Physics Today, 22 (1999); P.L. McEuen, Physics World, June 2000.

\bibitem{Tans97} S.J. Tans {\it et al.}, Nature {\bf 386}, 474 (1997).

\bibitem{Bockrath97} M. Bockrath {\it et al.}, Science {\bf 275}, 1922 (1997).

\bibitem{Liang2002} W. Liang, M.Bockrath, and H. Park, Phys. Rev. Lett. {\bf 88}, 126801 (2002).

\bibitem{Buitelaar2002} M. R. Buitelaar {\it et al.}, Phys. Rev. Lett. {\bf 88}, 156801 (2002).

\bibitem{Cobden2002} D.H. Cobden and J. Nygard, Phys. Rev. Lett. {\bf 89}, 046803 (2002).

\bibitem{Babic} B. Babic, T. Kontos, and C. Schonenberger, Phys. Rev. B {\bf 70} 235419 (2004)

\bibitem{Moriyama} S. Moriyama {\it et al.}, Phys. Rev. Lett. {\bf 94}, 186806  (2005).

\bibitem{Sapmaz} S. Sapmaz {\it et al.}, Phys. Rev. B {\bf 71}, 153402 (2005).

\bibitem{Jarillo-Herrero} P. Jarillo-Herrero {\it et al.}, Phys. Rev. Lett. {\bf 94}, 156802 (2005).

\bibitem{Zheng2002} B. Zheng {\it et al.}, Nano Lett. {\bf 2}, 895 (2002).

\bibitem{QDreview} L.P. Kouwenhoven, C.M. Marcus, P.L. McEuen, S. Tarucha, R.M. Westervelt, and N.S. Wingreen, in  Mesoscopic Electron Transport, ed. by L.P. Kouwenhoven, G. Schon, and L.L. Sohn, (Kluwer, 1997), p. 105.

\bibitem{ABG} I.L. Aleiner, P.W. Brouwer and L.I. Glazman, Physics Reports {\bf 358}, 309 (2002).

\bibitem{Oreg2000} Y. Oreg, K. Byczuk, and B.I. Halperin, Phys. Rev. Lett. {\bf 85}, 365 (2000).

\bibitem{Ke2003} S-H. Ke, H.U. Baranger, and W. Yang, Phys. Rev. Lett. {\bf 91}, 116803 (2003).

\bibitem{Saito} R. Saito,  G. Dresselhaus, and M. S. Dresselhaus, Physical Properties of Carbon Nanotubes, World Scientific, 1998. Points $\pm k_0$ originate from the equivalent points $K$ and $K'$ in the Brillouin zone of the graphite monolayer that makes the nanotube. We are illustrating the more common situation where $k_0 \neq 0$. For some metallic nanotubes $k_0 = 0$ in which case our considerations should also hold. 

\bibitem{Ando2000} T. Ando, Semicond. Sci. Technol. {\bf 15}, R13 (2000).

\bibitem{McEuen99} P.L. McEuen, {\it et al.}, Phys. Rev. Lett. {\bf 83}, 5098  (1999).

\bibitem{Kong1999} K. Kong, S. Han, and J. Ihm, Phys. Rev. B {\bf 60}, 6074 (1999).

\bibitem{Alhassid2000} Y. Alhassid, Rev. Mod. Phys. {\bf 72}, 895 (2000).

\bibitem{Averin1992} D.V. Averin and Yu.V. Nazarov, in Single Charge Tunneling: Coulomb Blockade Phenomena in Nanostructures, ed. by H. Grabert and M. H. Devoret (Plenum Press, 1992), p. 217.

\bibitem{Matveev2004} K. A. Matveev, Phys. Rev. Lett. {\bf 92}, 106801 (2004); Phys. Rev. B {\bf 70}, 245319 (2004). 

\bibitem{Kurland2000} I. L. Kurland, I. L. Aleiner, and B. L. Altshuler, Phys. Rev. B {\bf 62}, 14886 (2000).

\end{references}
\end{document}